# Circuit Design for Predictive Maintenance

**Taner Dosluoglu[1], Member, IEEE, Martin MacDonald[1]**
[1]Weeteq LTD, Glasgow, UK

Corresponding author: Taner Dosluoglu (e-mail: taner@5g3i.co.uk).

Weeteq Ltd. is part of 5G innovation and investment initiative (5G3i) based in Glasgow. 5G3i is a cohort company in Smart Things Accelerator Centre (STAC), "This work was supported in part by Scottish Enterprise and sponsors of STAC."

**ABSTRACT** Industry 4.0 has become a driver for the entire manufacturing industry. Smart systems have enabled 30% productivity increases and predictive maintenance has been demonstrated to provide a 50% reduction in machine downtime. So far, the solution has been based on data analytics which has resulted in a proliferation of sensing technologies and infrastructure for data acquisition, transmission and processing. At the core of factory operation and automation are circuits that control and power factory equipment, innovative circuit design has the potential to address many system integration challenges. We present a new circuit design approach based on circuit level artificial intelligence solutions, integrated within control and calibration functional blocks during circuit design, improving the predictability and adaptability of each component for predictive maintenance. This approach is envisioned to encourage the development of new EDA tools such as automatic digital shadow generation and product lifecycle models, that will help identification of circuit parameters that adequately define the operating conditions for dynamic prediction and fault detection. Integration of a supplementary artificial intelligence block within the control loop is considered for capturing non-linearities and gain/bandwidth constraints of the main controller and identifying changes in the operating conditions beyond the response of the controller. System integration topics are discussed regarding integration within OPC Unified Architecture and predictive maintenance interfaces, providing real-time updates to the digital shadow that help maintain an accurate, virtual replica model of the physical system.

## I. INTRODUCTION

Industry 4.0 and digital transformation offer significant benefits, such as 30% improvement in productivity and 50% reduction in machine downtime. Predictive maintenance plays a key role in delivering these benefits, introducing digital transformation for equipment maintenance and integration, to system level optimization and cost effective operations [1]. Currently implemented with a data centric approach based on the observation of systems, this has resulted in a proliferation of sensing and infrastructure investment to enable data acquisition, transmission, and processing [2]. A data-driven culture is an important part of improving operations efficiency, product quality, customer demand and service excellence in manufacturing. However, it is unnecessary and inefficient to approach factory floor equipment as a black box that needs to be observed. Data analytics should be complimented with components that seamlessly integrate with system level platforms. Smart components can be designed with embedded artificial intelligence that provide existing information autonomously, already stored in the correct format without any external data analytics. There is a need for a new design approach and disruption of the current data centric focused solutions.

## II. BACKGROUND

The semiconductor industry has been at the forefront of continuous improvement and scaling which has been going on for decades. While the process node dimensions capture the headlines, this has been made possible by the continuous improvement in the circuit design methodology which already includes extensive optimization for handling complex process corners, design for test, design for manufacturing (Design for everything in short) and EDA tools that handle multi-parameter optimization. During this optimization, circuit designers spend most of their time understanding how the circuit recovers from stress conditions and deviations due to process corners and how the circuit responds to unexpected operating conditions. The optimum circuit design is delivered based on the design constraints for testability and operation specifications. The knowledge and insights obtained during the design phase regarding stress conditions and failure modes are not captured in the same comprehensive manner as those are for testability of the device. This is a lost opportunity for better predictive maintenance where circuits can be essential components providing data regarding their operating environment. We envision a set of EDA tools for automatic digital shadow generation that captures all stress conditions and failure modes (significant system response deviation in general) as data constructs that are optimized for efficient handling by subsequent artificial intelligent processes. A model of the circuit as a digital shadow would be generated specifically for the purpose of handling data constructs transmitted during the operation of the device allowing accurate representation of the physical component to be made available for system level digital twin. Board design and manufacturing will include hierarchical digital shadow of the components and configuration for handling the interface between the sub-component digital shadows and the system level digital twin. Enabling technologies are already in place for innovation and disruption of the data centric approach in smart manufacturing. Cyber-Physical Systems allow virtual models to be deployed that represent the physical components and allow cognitive functions such as optimization and the decision making process to be more efficient based on the virtual replica (Digital Twin). Virtual sensors can be deployed in the system and reduce cost. Edge computing allows processing of the data closer to the physical device and reduce data transmission as well as latency.

All of these are steps in the right direction but they do not go far enough. In this paper, we present a new approach based on circuit level artificial intelligence solutions, integrated within control and calibration functional blocks during the design phase improving the predictability and adaptability of each component for predictive maintenance. This is envisioned as similar to the design for testability (DfT) approach where test structures are inserted during the design stage to measure testability and improve controllability and observability. Today, there are many tools available to designers to accommodate automatic scan chain insertion, test vector pattern generation, and determine test coverage. Analog designers can easily access guidelines for making architectural choices that reduce test time and cost by allowing observability through digital test multiplexers.

In the near future, similar tools will be available to enable design for predictive maintenance (PdM). These tools will enable predictability of the circuits based on the operating environment as well as their own aging process. Some of the key components will perform automatic digital shadow generation (ADSG), identification of key product lifecycle parameters, identification of circuit parameters that adequately define the operating conditions for dynamic prediction and fault detection.

## III. PROPOSED APPROACH

The consideration for predictability needs to be addressed during the design phase and there are many solutions that can be adapted without any changes to the controller architecture. We have chosen to change the architecture and include a supplementary controller in our solution. In our initial designs we have used control architectures with supplementary artificial intelligence within the closed-loop feedback that generates vectors based on deviation of the system response from an ideal response. In essence, the supplementary control block captures non-linearities and gain/bandwidth constraints of the main controller (designed for optimal control) and identifies changes in the operating conditions beyond the response of the controller. It allows simplification in the design where complex state transitions can be designed in the supplementary block. Configuration of the controller can be dynamic, based on the product lifecycle parameter vector. Anomalies based on operating conditions (or component faults) can be separated from system response constraints. The digital shadow can be maintained accurately for the entire product lifecycle with significantly less data compared to systems based on observation/sensors. The data reduction will be several orders of magnitude and have the biggest impact on the overall project cost for predictive maintenance initiatives. Reduction of latency has a similar impact on the scalability of distributed systems. These new circuits with improved adaptive response and quantified predictability coverage, will be key enablers for self-aware, system-aware components of cost effective system level optimization. At the system level, we anticipate wider adoption of explainable artificial intelligence and differential equation based machine learning that will take advantage of the real-time accurate digital shadow that will accompany these new components.

The modified architecture with supplementary artificial intelligence for improving predictability of the system is shown as a dashed line region in Figure 1. The supplementary unit monitors deviations of the system response from target value and improves the controller response for fast load transients, unexpected operating conditions and nonlinearities in the system. The integration of artificial intelligence within controller architecture has generated interest early on with pioneering efforts in the early 1990s [3]. The proposed supplementary control integration eliminates some of the hardware limitation in achieving ultra-low latency response by reframing the problem as an error correction reducing the difference between current value and the target value at the input of the modulator.

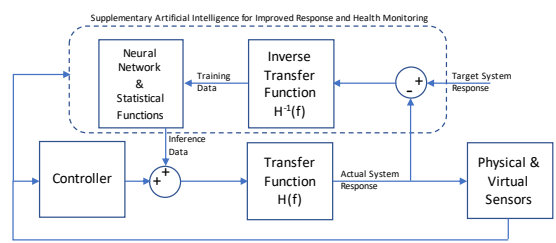

**FIGURE 1. Architecture of Circuit Design with Predictability.**

A predictive maintenance framework is built around use cases in the target application. A component that is part of the controller for motor drive is shown in Figure 2. The design phase includes the generation of digital shadow for the controller and integration with models for all components shown including the model for the motor.

The proposed approach provides real-time data from components and allows dynamic configuration for system level optimization. This dynamic optimization and real-time data access does not compromise security of the system compared to the current industrial IoT connected controllers and sensors. In fact, the automatic digital shadow will enable an additional level of security where product lifecycle and circuit parameter vectors can be used for encryption. Instead of allowing independent configuration of parameters to cover the entire operating conditions, the embedded control unit defines a consistent set of parameters for different operating conditions and ageing of the product.

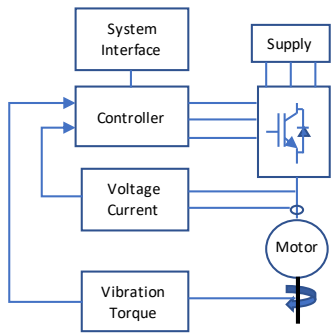

**FIGURE 2.** Motor Drive System.

## IV. SYSTEM INTEGRATION

The system integrators providing cloud-edge collaboration platforms will most likely be key enablers for the adoption of digital shadow as a requirement for components. They can provide models that define interface requirements and operational constraints that need to be validated. This will allow system integrators to provide their requirements to the component vendors in a format that can be part of the component design release process. In other words, system integrators would provide simulation models for component vendors to validate their design as ready/qualified for predictive maintenance in the corresponding system platforms. Key elements of this qualification are; optimum configuration for the part, and compatibility with the system platform for maintaining an accurate representation of the physical component for the entire product lifecycle.

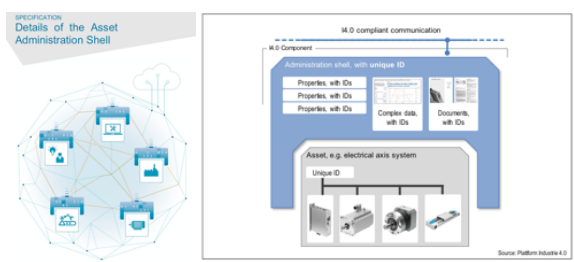

**FIGURE 3.** OPC UA Asset Administration Shell (Source: Platform Industrie 4.0_Specification).

In the Design for PdM approach, decision making processes within condition-based maintenance (CBM), Prognostics and Health Management (PHM) and Remaining Useful Life (RUL) methodologies are combined and taken into account during the design phase, where the parameters that have significance for predictive maintenance are identified and made available via the interface between the device and the system. These significant parameters are correlated to the process corners and operating condition simulations during design validation. The interface is based on the high level architecture for Cyber-Physical systems and integration of components that have been presented in numerous papers and addressed in standards.

Platform independent information modelling and access is already defined by OPC UA specification (figure 3)[4]. The architecture of the ultra-edge components and system interface is based on existing standards (figure 4). The asset includes integrated artificial intelligence within closed loop control and calibration and generate critical data ready to be used within the digital shadow generated during the design phase. Digital Shadow is included with the Asset Administration Shell and provides I4.0 Compliant interface

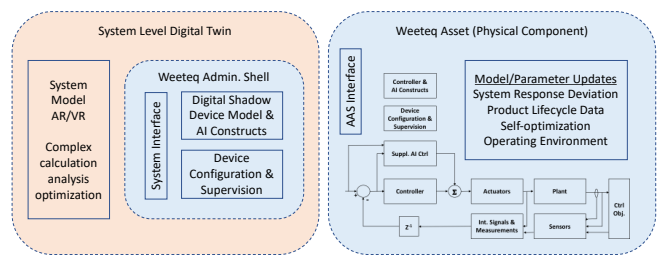

**FIGURE 4.** Weeteq Ultra-Edge and Digital Shadow.

## V. CONCLUSION

We have described circuit level artificial intelligence solutions that complement the control and calibration functions with semiconductor devices. These solutions provide two distinct benefits, namely, improve the system response of the device and provide system level interface predictive maintenance applications. System level improvements focus on nonlinear system response and bandwidth and power/gain limitations of the control loop. Predictive Maintenance interface provides real-time updates to the digital shadow and help maintain an accurate virtual replica model of the physical system.

The proposed circuit design approach has the potential to address many challenges faced by the adopters of smart technology. In order to achieve its full potential, it will need to be embraced by the entire ecosystem, including EDA vendors, semiconductor manufacturers, and smart manufacturing system platform providers.

## ACKNOWLEDGMENT

We would like to thank Smart Things Accelerator Centre (STAC), especially all the mentors and sponsors. Special thanks to Paul Wilson at STAC and Ronnie Darroch at Plexus for their guidance, support and valuable discussions.

## REFERENCES AND FOOTNOTES

[1] J. Yan, Y. Meng, L. Lu and L. Li, "Industrial Big Data in an Industry 4.0 Environment: Challenges, Schemes, and Applications for Predictive Maintenance," in IEEE Access, vol. 5, pp. 23484-23491, 2017, doi: 10.1109/ACCESS.2017.2765544.

[2] Y. Liu, W. Yu, T. Dillon, W. Rahayu and M. Li, "Empowering IoT Predictive Maintenance Solutions With AI: A Distributed System for Manufacturing Plant-Wide Monitoring," in IEEE Transactions on Industrial Informatics, vol. 18, no. 2, pp. 1345-1354, Feb. 2022, doi: 10.1109/TII.2021.3091774.

[3] S. Zhang, "Artificial Intelligence in Electric Machine Drives: Advances and Trends." arXiv preprint arXiv:2110.05403 (2021).

[4] OPC UA Companion Specification "Asset Administration Shell"